\documentclass[twocolumn,letterpaper,aps,prc,superscriptaddress,nofootinbib,showpacs,floatfix]{revtex4-1}
\usepackage{comment}
\usepackage{setspace}
\usepackage{color}
\usepackage{xcolor}
\usepackage{indentfirst}
\usepackage{xspace}
\usepackage{verbatim}
\usepackage{amssymb}
\usepackage{lipsum}
\usepackage{amsmath}
\usepackage{microtype}
\usepackage{booktabs}

\usepackage[export]{adjustbox}
\usepackage[T1]{fontenc}
\usepackage{float}
\usepackage{wrapfig}
\usepackage{graphicx}
\usepackage{placeins}

\usepackage{epsfig}

\usepackage{diffcoeff}
\usepackage{amsmath}
\usepackage{amssymb}
\usepackage{amsthm}
\usepackage{array}
\usepackage{tabularx}
\usepackage[utf8]{inputenc}
\graphicspath{ {figures/} }
\usepackage{lineno}
{\large }

\usepackage{natbib}
\setcitestyle{square, comma, numbers,sort \& compress, super}
\usepackage{appendix}
\usepackage{dcolumn}
\usepackage{bm}
\usepackage{hyperref}
\hypersetup{
    colorlinks=true,
    linkcolor=black,
    filecolor=magenta,      
    urlcolor=blue,
    citecolor=black,
    pdftitle={Overleaf Example},
    pdfpagemode=FullScreen,
    }
\usepackage[a4paper]{geometry}
\usepackage{ragged2e}
\begin{document}
\title{Monte-Carlo Study Of Higher-Order Cumulants of Net-Particle Distributions in \(p+p\) Collisions at \(\sqrt{s}\) = 13 TeV}

	\author{Abdussamad M}
	\affiliation{Department of Physics, Indian Institute of Technology Bombay, Mumbai, India-400076}
	\author{Rahul Verma }
	\affiliation{Department of Physics, Indian Institute of Technology Bombay, Mumbai, India-400076}
    \author{Nirbhay Kumar Behera}
	\email{Corresponding Author: nirbhaykumar@cutn.ac.in}
	\affiliation{Department of Physics, Schools of Basic and Applied Sciences, Central University of Tamil Nadu, Thiruvarur, India-610005}
    \author{Sadhana Dash}
	\affiliation{Department of Physics, Indian Institute of Technology Bombay, Mumbai, India-400076}
    \author{Basanta Kumar Nandi}
	\affiliation{Department of Physics, Indian Institute of Technology Bombay, Mumbai, India-400076}

\begin{abstract}
 Measurement of higher order cumulants of the distributions of conserved quantities, like net-charge, net-baryon and net-strangeness in heavy-ion collisions, is proposed as a sensitive tool to determine the freeze-out parameters and the nature of phase transitions at the LHC energies. Baseline measurements for heavy-ion collisions are essential to understand the experimental measurements. Recently, several experimental observations have shown some QGP-like scenarios in small systems (pp collisions). We report the first Monte-Carlo study of the measurements of cumulants and their ratios for net-charge, net-hadron, net-kaon, net-baryon, and net-proton distributions in pp collisions at $\sqrt{s}$=13 TeV using pQCD models like Pythia8 and Herwig. We also discuss the effect of different particle production mechanisms on the higher-order cumulants. This simulation study will serve as a baseline for future measurements at the LHC. Furthermore, it will shed more light on the measurement of cumulants and the connection between small systems and heavy-ion collisions at the LHC.

\end{abstract}
\maketitle
\section{Introduction} \label{introduction}

Heavy-ion collision experiments aim to explore the Quantum Chromodynamics (QCD) phase diagram in the Temperature (T) and baryo-chemical potential ($\mu_B$) plane. The study of higher-order cumulants of conserved quantities like net-charge, net-baryon, and net strangeness and their ratios are proposed as powerful tools to explore the QCD phase diagram, such as the critical point and freeze-out parameters in a model-independent way \cite{Stephanov:1998dy, Stephanov:1999zu, Aoki:2006we, Stephanov:2008qz, Friman:2011pf}. The Beam Energy Scan (BES) Program of Relativistic Heavy-ion Collision (RHIC) experiment at Brookhaven National Laboratory (BNL) aims to explore the QCD phase diagram at higher $\mu_B$ values. Recently, some interesting experimental results of higher order cumulants of net-proton number distributions are reported by the STAR experiment, which will be helpful to locate the QCD critical point \cite{STAR:2013gus, STAR:2020tga, STAR:2021iop}. On the other hand, the heavy-ion program at the Large Hadron Collider (LHC) experiment at CERN will explore the QCD phase diagram at $\mu_B \approx 0$.  

\par
Lattice QCD (LQCD) calculations at vanishing baryonic chemical potential (\( \mu_B \approx 0\)) and large temperatures have established that the nature of the quark-hadron phase transition is a smooth crossover \cite{Aoki:2006we}. In a thermal system where the volume remains constant, fluctuations in conserved quantities are related to thermodynamic susceptibilities \cite{Karsch:2010ck}. These susceptibilities can be calculated within the framework of LQCD. The susceptibilities are defined as the partial derivatives of the reduced pressure with respect to the reduced chemical potential as given below \cite{Karsch:2010ck}.
\[
   \chi_n^{N=B,S,Q}=\frac {\partial^n (p/T^4)}{\partial(\mu_{N}/T)^n}
\]
Here, \(B\), \(S\), and \(Q\) represent the baryon number, strangeness, and electric charge, respectively. One can relate the cumulants of the conserved charge multiplicity distribution to the corresponding quantum-number susceptibilities as follows, \(C_n^N = \frac{V}{T^3}\chi_n^N\) \cite{Karsch:2010ck}. The variables \(p\), \(V\), and \(T\) signify the pressure, volume, and temperature of the system. \(C_n\) denotes the \(n^{th}\) order cumulants of multiplicity distributions. In experimental measurements, cumulant ratios are reported to mitigate the volume effects. So, the ratios of cumulants are directly connected to the ratios of various order susceptibilities. That allows for establishing a direct connection between the experimental results and LQCD. It is shown that the freeze-out parameters (T and $\mu_B$) can be estimated from the ratio of cumulants \cite{Friman:2011pf, Karsch:2010ck}. Recent LQCD calculation and statistical model estimation of freeze-out temperature shows that it is close to the crossover temperature \cite{Bhattacharya:2014ara, Stachel:2013zma}, which can be constrained by the experimental measurements at the LHC. ALICE experiment at the LHC has reported the second cumulant of net-proton number distributions, and preliminary results of it up to fourth order relatively in small kinematic window \cite{Behera:2018wqk, ALICE:2019nbs}. The upcoming precision measurement of higher order cumulants using high statistics Pb-Pb collision data at the LHC will shed more light on estimating the freeze-out temperature in a model-independent way \cite{Citron:2018lsq}.

\par
Meanwhile, some recent experimental results suggest the possible formation of QGP medium in high-multiplicity pp collisions \cite{CMS:2010ifv, ALICE:2016fzo, CMS:2016fnw}. This has drawn much attention to understanding the QGP-like medium in small systems \cite{OrtizVelasquez:2013ofg, Blok:2017pui}. Therefore, experimental measurement of cumulants of conserved quantities in high-multiplicity pp collisions at the LHC energies will help understand the nature of the phase transition. In this regard, the study of the higher order cumulants using QCD-based models, like Pythia8 \cite{Sjostrand:2014zea, Bierlich:2022pfr} and Herwig \cite{Webber:1983if, Bahr:2008pv} may help to comprehend the effect of various hadronisation process on them. These model studies can be used as a baseline for the ongoing or future experimental measurements in pp collisions. The collision energy dependence of cumulant ratios (\(S\sigma\) and \(k\sigma^2\)) were measured in both \(p+p\) and \(Au+Au\) collisions \cite{STAR:2013gus}. However, the large uncertainties arising from limited statistics impeded the extraction of definitive conclusion. 
\par
It is crucial to establish the consistency of higher-order cumulant ratios with expectations from a non-critical baseline, indicating the absence of critical behaviour. One often uses the Skellam distribution (arising from the difference of two variables following independent Poisson distributions)
In this work, we present the event multiplicity-dependent Monte-Carlo study of the measurements of the higher-order cumulants of the conserved quantities, like net-charge, net-baryon and net-strangeness, using pQCD inspired models like Pythia8 and Herwig. A systematic study has been carried out by considering different options for parton-level interactions available in these models as a baseline to study the net-particle multiplicity fluctuations.  

\section{Observables}
Cumulants express the characteristics of a distribution. Up to the sixth order, the cumulants of a distribution are defined as follows:
\begin{align*}
C_1 &= \langle N \rangle \\
C_2 &= \langle (\delta N)^2 \rangle \\
C_3 &= \langle (\delta N)^3 \rangle \\
C_4 &= \langle (\delta N)^4 \rangle - 3\langle (\delta N)^2 \rangle^2 \\
C_5 &= \langle (\delta N)^5 \rangle - 10\langle (\delta N)^2 \rangle\langle (\delta N)^3 \rangle \\
C_6 &= \langle (\delta N)^6 \rangle + 30\langle (\delta N)^2 \rangle^3 - 15\langle (\delta N)^2 \rangle\langle (\delta N)^4 \rangle \\
& \quad- 10\langle (\delta N)^3 \rangle^2
\end{align*}
where N represents the event-by-event conserved quantity and $\delta N = N - \langle N \rangle$. The symbol $\langle N \rangle$ represents the events' average value of $N$. Here, the first ($C_1$) and second ($C_2$) order cumulants are the mean ($M$) and variance ($\sigma^2$) of the distribution. Usually, one takes the following appropriate ratios of
these cumulants:
\begin{align*}
\frac{C_3}{C_2} = \frac{\chi_{(3)}^{N}}{\chi_{(2)}^{N}} = S\sigma, \ \ \ \ \ \frac{C_4}{C_2} = \frac{\chi_{(4)}^{N}}{\chi_{(2)}^{N}} = \kappa\sigma^2
\end{align*}
Here, $S$ and $\kappa$ represent skewness and kurtosis, respectively, and $\chi^N$ denotes the corresponding susceptibility. With the given definitions, one can determine various cumulants and their ratios from the obtained event-by-event net-particle multiplicity distributions. 
\par

\section{Analysis}
\subsection{Monte Carlo Sample}
This study uses Monte Carlo data of \(p+p\) collisions at  \( \sqrt{s} \)= 13 TeV created with Pythia8 and Herwig models. Overall, close to 500 million events were analyzed. The input parameters used for event generation for Pythia8 and Herwig are listed in Table \ref{tab:pythia-parameters} and Table \ref{tab:herwig-parameters}, respectively  \cite{Sjostrand:2014zea, Bierlich:2022pfr, Bahr:2008pv}. 
\begin{table}[htbp]
    \centering
    \begin{tabular}{@{}ll@{}}
        \toprule
        \textbf{Parameter} & \textbf{Value} \\ \midrule
        Version & Pythia 8.3.10 \\
        Tune & Monash 2013 \\
        Color Reconnection & On/Off \\
        ColourReconnection:mode & 0 \\
        BeamRemnants:remnantMode & 1 \\
        PartonVertex:setVertex & On \\
        Multiparticle Interaction & On \\ 
        MultipartonInteractions:pT0Ref & 2.659 \\ \bottomrule
    \end{tabular}
    \caption{Pythia8 Parameters}
    \label{tab:pythia-parameters}
\end{table}

\begin{table}[htbp]
    \centering
    \begin{tabular}{@{}ll@{}}
        \toprule
        \textbf{Parameter} & \textbf{Value} \\ \midrule
        Version & Herwig 7.2.3 \\
        Tune & H7.1-Default \\ 
        Color Reconnection & On/Off \\
        ReconnectionProbability & 0.652710 \\ 
        colourDisrupt & 0.75 \\ 
        InvRadius & 1.489997 \\ 
        Multiparticle Interaction & On \\
        IdenticalToUE & 0 \\
        DLmode & 2 \\ \bottomrule
    \end{tabular}
    \caption{Herwig Parameters}
    \label{tab:herwig-parameters}
\end{table}
The difference between how Pythia8 and Herwig handle hadronisation is quite significant. Hadronisation is the non-perturbative mechanism to convert partons into hadrons at the typical hadronic scale. In Herwig, this process is accomplished through the cluster model, which is in the concept of colour pre-confinement inherent in the angular-ordered parton shower \cite{Webber:1983if,Amati:1979fg}. On the other hand, Pythia8 uses the Lund String model for hadronisation \cite{Andersson:1983jt}. The Lund String Model is based on approximating the confining colour field between a quark-antiquark pair with a massless relativistic string. Analysing higher-order cumulants using these two models provides insights into the underlying physics dynamics.

\subsection{Methodology}
Events in this analysis are classified into different multiplicity classes based on charged-particle multiplicity, obtained in the acceptance range of pseudorapidity (\(\eta\)) intervals of \(2.8 < \eta < 5.1\) and \(-3.7 < \eta  < -1.7\). The net-particle analysis considers various acceptance ranges for \(\eta\) and transverse momentum (\(p_T\)). For the analysis of net-charge and net-hadron, the chosen \(\eta\) range is \(|\eta| < 0.8\), and the \(p_T\)  interval is \(0.15 \leq p_{T} \leq 3.0\) GeV/c. For net-kaon, net-baryon, and net-proton, the \(\eta\) range remains \(|\eta| \leq 0.8\), with a \(p_T\) range of \(0.2 \leq p_{T} \leq 2.0\) GeV/c. 
\par
We compare the results obtained from the Pythia8 and Herwig models, along with two specific variations: Colour Reconnection (CR) enabled (On) and disabled (Off), and the influence of MBWC methods in both models. The primary motivation for these comparisons is to enhance our understanding of the particle production mechanism by investigating how parameter variations influence the model predictions. Secondly, we aim to assess whether conventional physics mechanisms can accurately explain the observed data patterns in \(p+p\) collisions without invoking the creation of a deconfined medium.
\par
We have taken varied approaches to net-particle analysis. The inclusive method included events where both the number of particles and anti-particles were found to be zero in the acceptance. Generally, the experimental data analysis considers the inclusive method. In contrast, we have also considered the exclusive method, which excludes events where both the number of particles and anti-particles are found to be zero. The events were excluded in view of the specific observations where no particle-antiparticle pairs existed in the considered kinematic acceptance while they were present in the phase space outside the analysis acceptance. This comparison between exclusive and inclusive methods yields valuable insights into the analytical approach.

Cumulants of net-particles are known to be artificially enhanced in experimental measurements due to initial volume fluctuation \cite{Luo:2013bmi}in heavy ion collisions. To mitigate this effect, one approach is the application of the Multiplicity Bin Width Correction method (MBWC) \cite{Luo:2013bmi}. Using this method, the cumulant, $C_n$ for a wider multiplicity bin is estimated as follows:
\begin{align*}
C_n = \sum_{r} \omega_r C_{(n,r)} \ \ , \ \quad \omega_r = \frac{N_r}{\sum_{r} N_r} 
\end{align*}
where \(N_r\)  and  \( C_{(n,r)} \) represent the number of events and the \(n^{th}\) order cumulants in the \(r^{th}\) multiplicity bins, respectively. In this study, we have considered the charged particle multiplicity in terms of percentile bins each of 1$\%$ width. The final results are presented at 20$\%$ bin width using MBWC.
\par
The subgroup method has been used to calculate statistical uncertainties. The subgroup method involves dividing the initial data sample into \(S\) subgroups randomly \cite{ALICE:2017jsh}. The cumulants (\(C_n\)) for a given multiplicity percentile bin is computed as, 
\[
C_n = \frac{\sum_{i=1}^S{C_n^i}}{S},
\]
where \(S\) represents the number of subgroups and \(C_{n}^i\) pertains to the cumulants computed for each subgroup. Then the  statistical error (\(\delta C_n\)) for a given multiplicity percentile bin is determined by 
\[
\delta C_n = \frac{\sigma}{\sqrt{S}} \ \ , \ \ \mathrm{where}\hspace{0.1 in} \sigma =\sqrt{\frac{\sum_{i=1}^{S}(C_n^i-<C_n>)^2}{S-1}} 
\]
By using the MBWC, the statistical uncertainty of \(C_n\) for wider multiplicity percentile bin is estimated as follows. 
\[
\epsilon_n=\left(\frac{\sum_{i=1}^{k} m_i^2 \delta C_{n,i}^2}{\sum_{i=1}^{k}m_i^2}\right)^{1/2}
\label{eq 3.23}
\]
Where $m_i$ denotes the number of events in the $i^{th}$ multiplicity bin, and $\delta C_{n,i}$ represent the $n^{th}$ order cumulant error of the $i^{th}$ multiplicity percentile bin. Here, $k$ represents the total number of multiplicity percentile bins.

\section{Result and Discussions}

\begin{figure*}
	\centering 
	\includegraphics[width=0.99\textwidth]{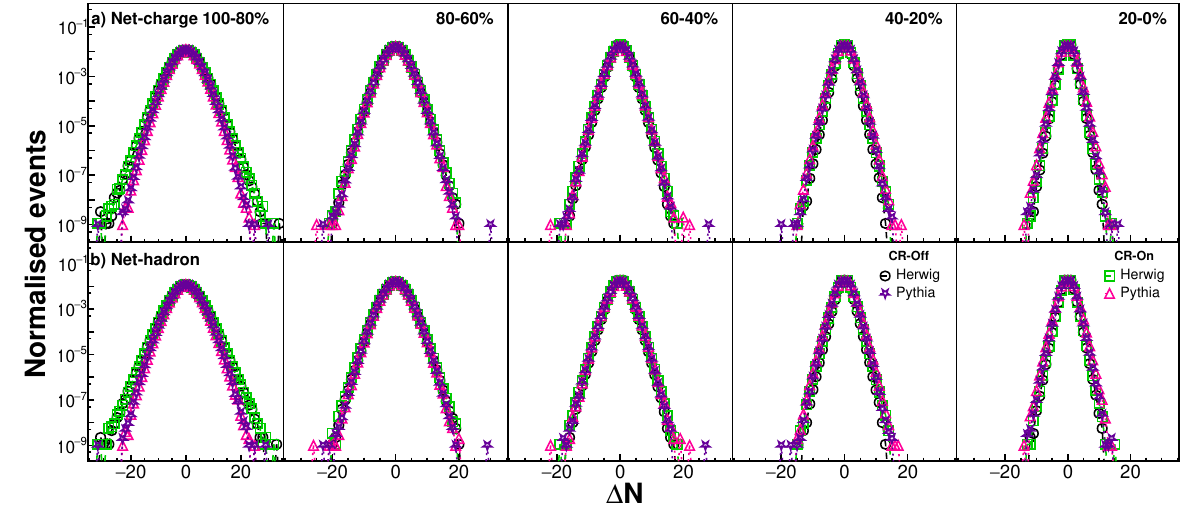}	
	 \caption{Normalized multiplicity distributions of (a) net-charge and (b) net-hadron in \(p+p\) collisions at \( \sqrt{s}\) = 13 TeV in five multiplicity bins (0-20\%, 20-40\%, 40-60\%, 60-80\%, 80-100\%) obtained from Pythia8 and Herwig under CR-off and CR-on modes.}
	\label{netCH CRoff vs herwig}%
\end{figure*}

\begin{figure*}
	\centering 
	\includegraphics[width=0.99\textwidth]{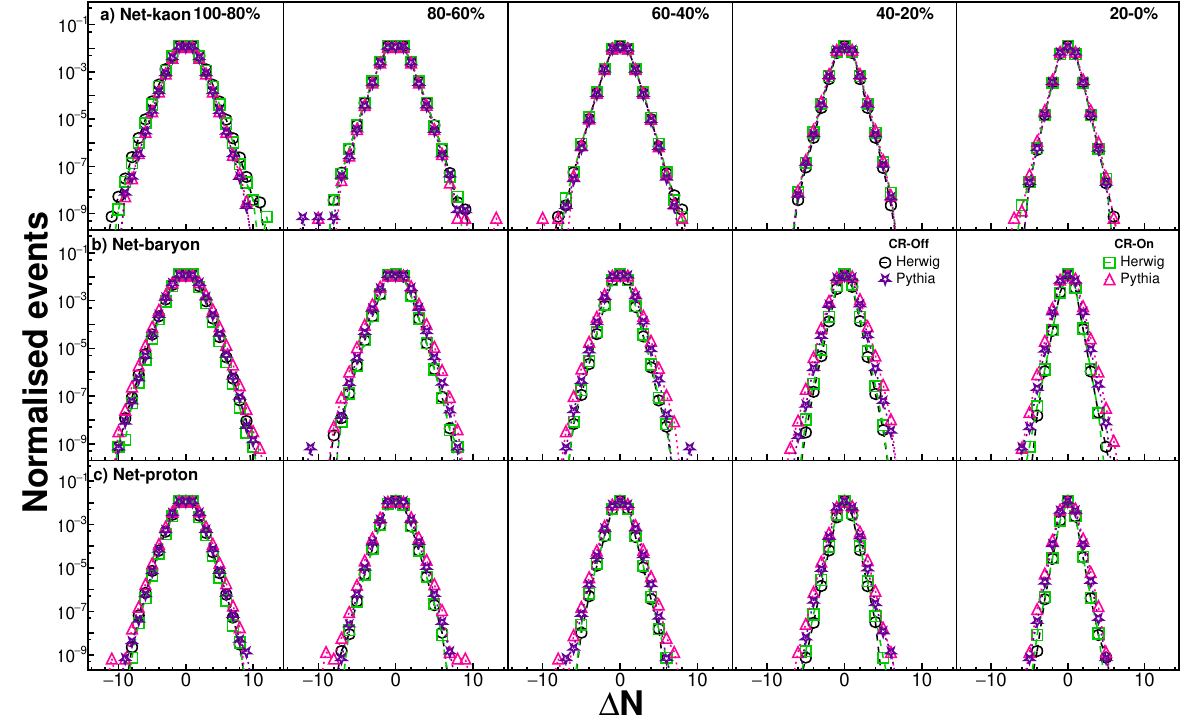}	
	\caption{Normalized multiplicity distributions of (a) net-kaon, (b) net-baryon, and (c) net-proton under CR-off and CR-on modes in \(p+p\) collisions at \(\sqrt s\) = 13 TeV across five multiplicity bins (0-20\%, 20-40\%, 40-60\%, 60-80\%, 80-100\%) with CR-off and CR-on obtained from Pythia8 and Herwig.} 
	\label{netKBP CRoff vs herwig}%
\end{figure*}

The normalized multiplicity distributions of net-charge and net-hadron for five different multiplicity bins are illustrated in Figure \ref{netCH CRoff vs herwig}. Similarly, the normalized multiplicity distributions of net-kaon, net-baryon, and net-proton for five different multiplicity bins are illustrated in Figure \ref{netKBP CRoff vs herwig}. The results presented in Figure \ref{netCH CRoff vs herwig} and \ref{netKBP CRoff vs herwig} are obtained with the modes CR-off and CR-on for both Pythia8 and Herwig models. The analysis reveals a consistent pattern in the net-charge and net-hadron multiplicity distributions across the entire range of multiplicity classes, except for the higher multiplicity bin, where Herwig exhibits a broader distribution than Pythia8 in both net-charge and net-hadron distributions. As the multiplicity decreases, the width of the distribution also decreases consistently for net-kaon, net-baryon, and net-proton for both models. No significant difference is observed between CR-off and CR-on modes for the distributions. 

\par

\begin{figure*}
\centering 	\includegraphics[width=0.99\textwidth]{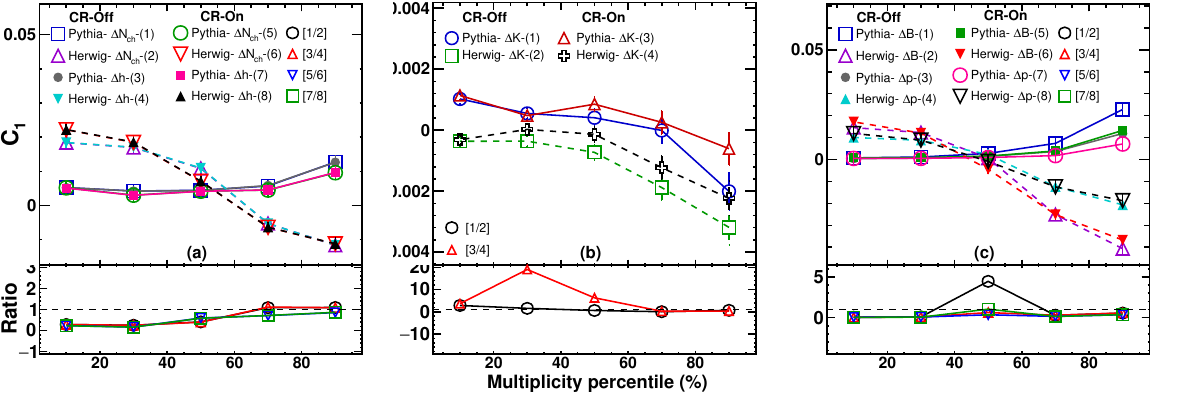}	
	\caption{Multiplicity-dependent first-order cumulant (\(C_1\)) of net-charge and hadron (a), net-kaon (b), and net-baryon and proton (c) distributions in \(p+p\) collisions at \( \sqrt s\) = 13 TeV.  Results are presented for the 20\% multiplicity bins range. Solid lines depict Pythia8 results, and dotted lines represent Herwig results. Vertical lines denote statistical uncertainties. The ratio plot illustrates the relative difference between Pythia8 and Herwig results.} 
	\label{PH_C1}%
\end{figure*}

\begin{figure*}
	\centering 
	\includegraphics[width=0.99\textwidth]{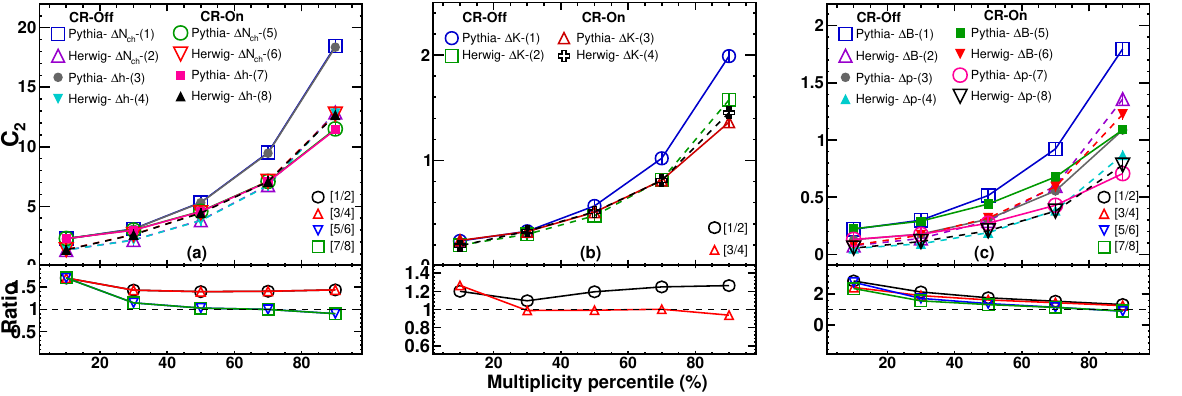}	
	\caption{Multiplicity-dependent second-order cumulant (\(C_2\)) of net-charge and hadron (a), net-kaon (b), and net-baryon and proton (c) distributions in \(p+p\) collisions at \( \sqrt s\) = 13 TeV.  Results are presented for the 20\% multiplicity bins range. Solid lines depict Pythia8 results, and dotted lines represent Herwig results. Vertical lines denote statistical uncertainties. The ratio plot illustrates the relative difference between Pythia8 and Herwig results.} 
	\label{PH_C2}%
\end{figure*}

\begin{figure*}
	\centering 
	\includegraphics[width=0.99\textwidth]{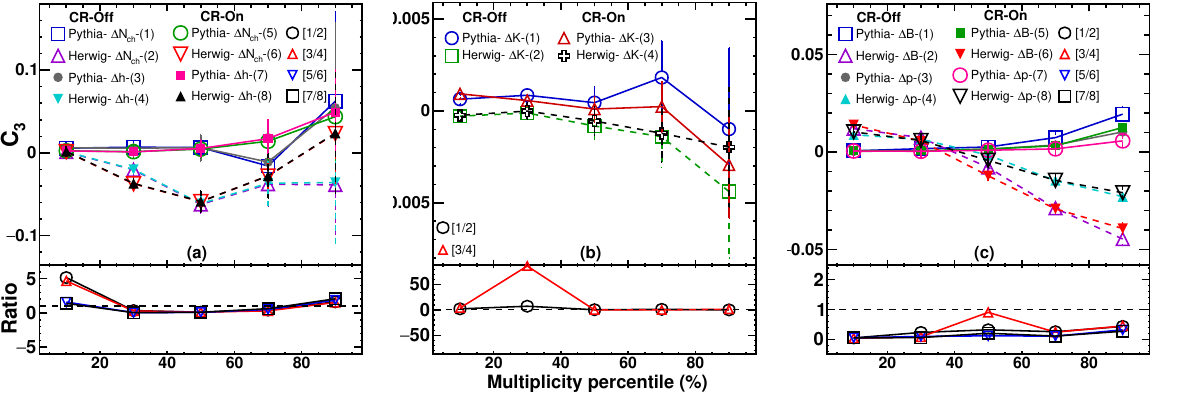}	
	\caption{Multiplicity-dependent third-order cumulant (\(C_3\)) of net-charge and hadron (a), net-kaon (b), and net-baryon and proton (c) distributions in \(p+p\) collisions at \( \sqrt s\) = 13 TeV.  Results are presented for the 20\% multiplicity bins range. Solid lines depict Pythia8 results, and dotted lines represent Herwig results. Vertical lines denote statistical uncertainties. The ratio plot illustrates the relative difference between Pythia8 and Herwig results.} 
	\label{PH_C3}%
\end{figure*}

\begin{figure*}
	\centering 
	\includegraphics[width=0.99\textwidth]{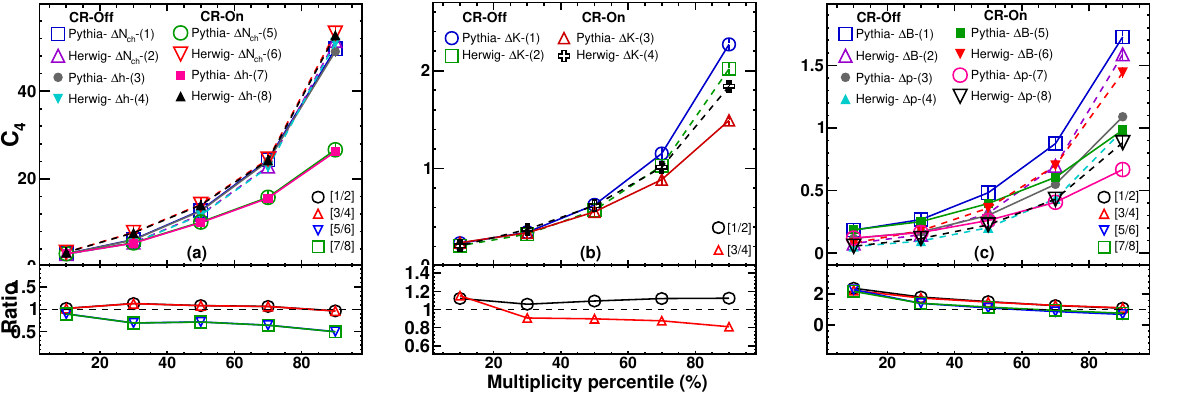}	
	\caption{Multiplicity-dependent fourth-order cumulant (\(C_4\)) of net-charge and hadron (a), net-kaon (b), and net-baryon and proton (c) distributions in \(p+p\) collisions at \( \sqrt s\) = 13 TeV.  Results are presented for the 20\% multiplicity bins range. Solid lines depict Pythia8 results, and dotted lines represent Herwig results. Vertical lines denote statistical uncertainties. The ratio plot illustrates the relative difference between Pythia8 and Herwig results.} 
	\label{PH_C4}%
\end{figure*}

\begin{figure*}
	\centering 
	\includegraphics[width=0.99\textwidth]{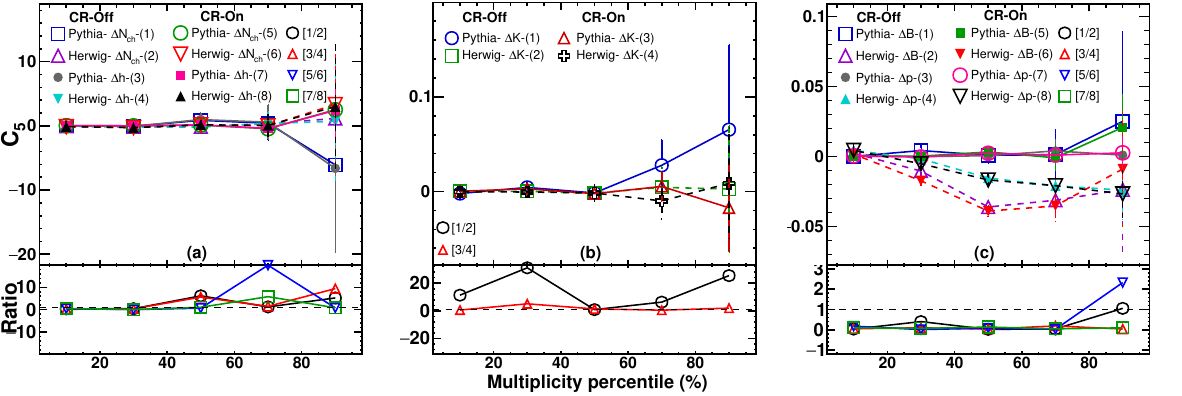}	
	\caption{Multiplicity-dependent fifth-order cumulant (\(C_5\)) of net-charge and hadron (a), net-kaon (b), and net-baryon and proton (c) distributions in \(p+p\) collisions at \( \sqrt s\) = 13 TeV.  Results are presented for the 20\% multiplicity bins range. Solid lines depict Pythia8 results, and dotted lines represent Herwig results. Vertical lines denote statistical uncertainties. The ratio plot illustrates the relative difference between Pythia8 and Herwig results.} 
	\label{PH_C5}%
\end{figure*}

\begin{figure*}
	\centering 
	\includegraphics[width=0.99\textwidth]{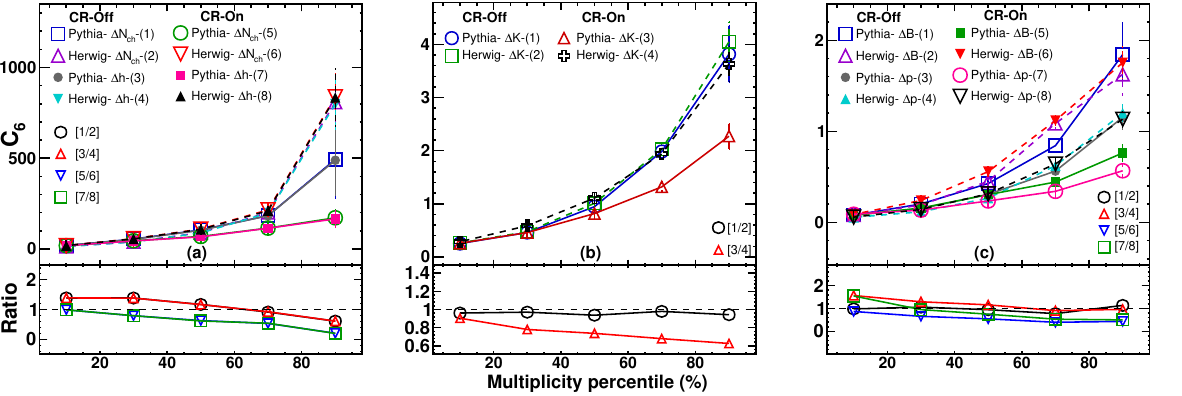}	
	\caption{Multiplicity-dependent sixth-order cumulant (\(C_6\)) of net-charge and hadron (a), net-kaon (b), and net-baryon and proton (c) distributions in \(p+p\) collisions at \( \sqrt s\) = 13 TeV.  Results are presented for the 20\% multiplicity bins range. Solid lines depict Pythia8 results, and dotted lines represent Herwig results. Vertical lines denote statistical uncertainties. The ratio plot illustrates the relative difference between Pythia8 and Herwig results.} 
	\label{PH_C6}%
\end{figure*}

\begin{figure*}
	\centering 
	\includegraphics[width=0.99\textwidth]{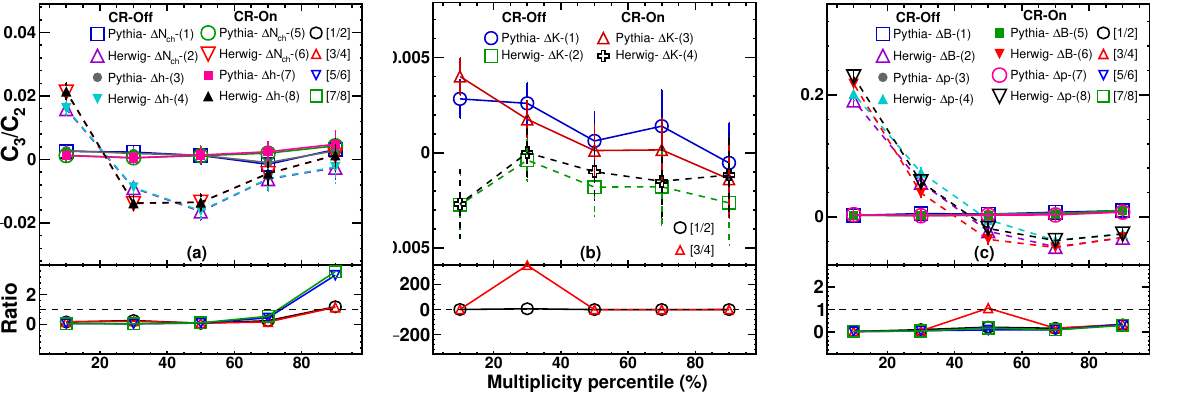}	
	\caption{Multiplicity-dependent cumulant ratio (\(C_3/C_2\)) of net-charge and hadron (a), net-kaon (b), and net-baryon and proton (c) distributions in \(p+p\) collisions at \( \sqrt s\) = 13 TeV.  Results are presented for the 20\% multiplicity bins range. Solid lines depict Pythia8 results, and dotted lines represent Herwig results. Vertical lines denote statistical uncertainties. The ratio plot illustrates the relative difference between Pythia8 and Herwig results.} 
	\label{PH_C32}%
\end{figure*}

\begin{figure*}
	\centering 
	\includegraphics[width=0.99\textwidth]{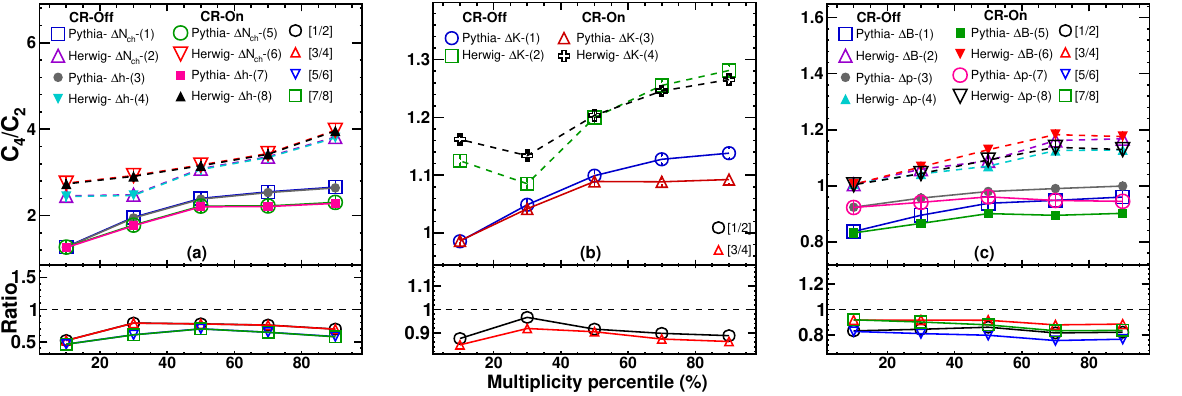}	
	\caption{Multiplicity-dependent cumulant ratio (\(C_4/C_2\)) of net-charge and hadron (a), net-kaon (b), and net-baryon and proton (c) distributions in \(p+p\) collisions at \( \sqrt s\) = 13 TeV.  Results are presented for the 20\% multiplicity bins range. Solid lines depict Pythia8 results, and dotted lines represent Herwig results. Vertical lines denote statistical uncertainties. The ratio plot illustrates the relative difference between Pythia8 and Herwig results.} 
	\label{PH_C42}%
\end{figure*}

\begin{figure*}
	\centering 
	\includegraphics[width=0.99\textwidth]{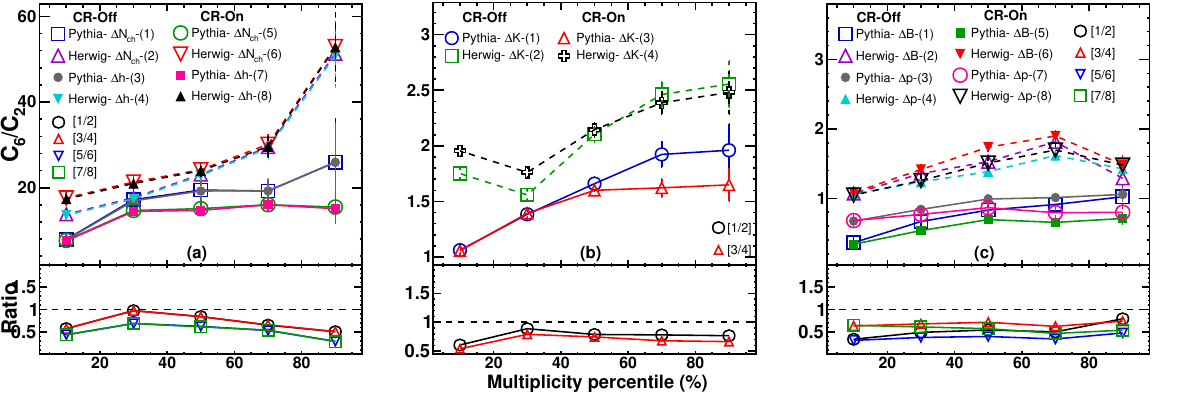}	
	\caption{Multiplicity-dependent cumulant ratio (\(C_6/C_2\)) of net-charge and hadron (a), net-kaon (b), and net-baryon and proton (c) distributions in \(p+p\) collisions at \( \sqrt s\) = 13 TeV.  Results are presented for the 20\% multiplicity bins range. Solid lines depict Pythia8 results, and dotted lines represent Herwig results. Vertical lines denote statistical uncertainties. The ratio plot illustrates the relative difference between Pythia8 and Herwig results.} 
	\label{PH_C62}%
\end{figure*}

Figures \ref{PH_C1}, \ref{PH_C2}, \ref{PH_C3}, \ref{PH_C4}, \ref{PH_C5}, and \ref{PH_C6} depict the multiplicity dependence of cumulants up to the sixth order of net-particles. Figure \ref{PH_C32}, \ref{PH_C42}, and \ref{PH_C62} illustrate the multiplicity dependence of cumulant ratios. The lower panel of these figures (Figure  \ref{PH_C1}, \ref{PH_C2}, \ref{PH_C3}, \ref{PH_C4}, \ref{PH_C5}, and \ref{PH_C6}) represents the ratio between the cumulants obtained from Pythia8 and Herwig. It can be seen from the ratios that there are differences between the cumulants values obtained from Herwig and Pythia8, and the difference is significant at high multiplicity event classes. The mean values of the net-charge, net-hadron, net-kaon, net-baryon, and net-proton distributions are close to zero, with minimal differences observed between CR-off and CR-on modes. The \(C_1\) value increases from CR-off to CR-on mode for a given model, with this effect being more pronounced in high multiplicity event classes. For net-charge, net-hadron, net-baryon, and net-proton distributions, the \(C_1\)  values increase with multiplicity in the Pythia8 model and decrease in the Herwig model. In the case of the net-kaon distribution, \(C_1\) decreases with increasing multiplicity in both models. \(C_2\) value of all distributions show an increasing trend with multiplicity, with a decline from CR-off to CR-on mode condition. The third cumulant,  \(C_3\), oscillates around zero for net-charge and net-hadron, with CR-off and CR-on values closely aligned in Herwig, while in Pythia8, CR-off increases slightly with multiplicity. \(C_3\) of net-kaon, both CR-off and CR-on decrease as multiplicity increases, with CR-on values slightly higher than CR-off. \(C_3\) results of net-baryon and net-proton obtained from Herwig show decreasing values with increasing multiplicity, whereas Pythia8 exhibits a slight increasing trend. Moreover, the $C_3$ value of net-baryon and net-proton for high multiplicity classes differs from each other. The fourth cumulant, \(C_4\),  values across all distributions rise with multiplicity, and the value decreases from CR-off to CR-on mode. There is no such significant difference observed for net-charge and net-hadron $C_4$ values. However, there is a clear difference of $C_4$ value is observed between net-baryon and net-proton. $C_5$ for net-hadron and net-charge distributions obtained from both models do not show any multiplicity dependence within the statistical uncertainties. \(C_5\) of net-kaon distributions obtained from both the models with CR-on and CR-off mode is close to zero within the statistical uncertainties. \(C_5\) of net-baryon and net-proton distributions are almost close to each other for a given model, and the estimation from Pythia8 is slightly higher than Herwig. \(C_6\) values for all distributions show an increasing trend with increasing multiplicity, and there is a decrease in the value from CR-off to CR-on mode. As observed in $C_4$ results, there is a significant difference between the $C_6$ results of net-baryon and net-proton for both models.

Figure \ref{PH_C32} illustrates \(C_3/C_2\) results of net-hadron, net-charge, net-kaon, net-baryon and net-proton. The \(C_3/C_2\) results of net-charge and net-hadron obtained from Pythia8 do not show any multiplicity dependence, however Herwig results oscillates from positive to negative values. For net-kaon, there is an increasing trend from CR-off to CR-on in both models. \(C_3/C_2\) of net-baryon and net-proton obtained from Herwig decreases with an increase in multiplicity, while the estimations from Pythia8 with CR-off and CR-on mode almost remain constant with multiplicity. 

\(C_4/C_2\) results of net-hadron, net-charge, net-kaon, net-baryon and net-proton are illustrated in Figure \ref{PH_C42}. \(C_4/C_2\) of net-charge and net-hadron show an increasing trend with increasing event multiplicity. In Pythia8, CR-off values are slightly higher than CR-on, while in Herwig, it is the reverse. For net-kaon, \(C_4/C_2\) increases slightly while going from low to high multiplicity event. In Pythia8, CR-off values are slightly higher than CR-on, whereas in Herwig, the values increase from CR-on to CR-off. Similarly, for net-baryon and net-proton, \(C_4/C_2\) values show a weak increasing trend with event multiplicity. Pythia8 shows a slightly higher \(C_4/C_2\) value for CR-off than CR-on mode, which is reversed in the case of the results obtained from the Herwig model. It can be seen that there are some differences between the \(C_4/C_2\) value of net-proton and net-baryon. 

Figure \ref{PH_C62} illustrates \(C_6/C_2\) results of net-hadron, net-charge, net-kaon, net-baryon and net-proton. \(C_6/C_2\) value of net-charge and net-hadron increases with increase in multiplicity. In Pythia8, CR-off values are slightly higher than CR-on, while in Herwig, both modes have values very close to each other. For net-kaon, \(C_6/C_2\) behaves almost similalry as its \(C_6/C_2\) results. Similarly, for net-baryon and net-proton, \(C_6/C_2\) values show a very weak increasing trend with event multiplicity, with Pythia8 showing a slightly higher value of \(C_6/C_2\) for CR-off than CR-on, which is reverse for Herwig.


\par
\begin{figure}
	\centering 
	\includegraphics[width=0.49\textwidth]{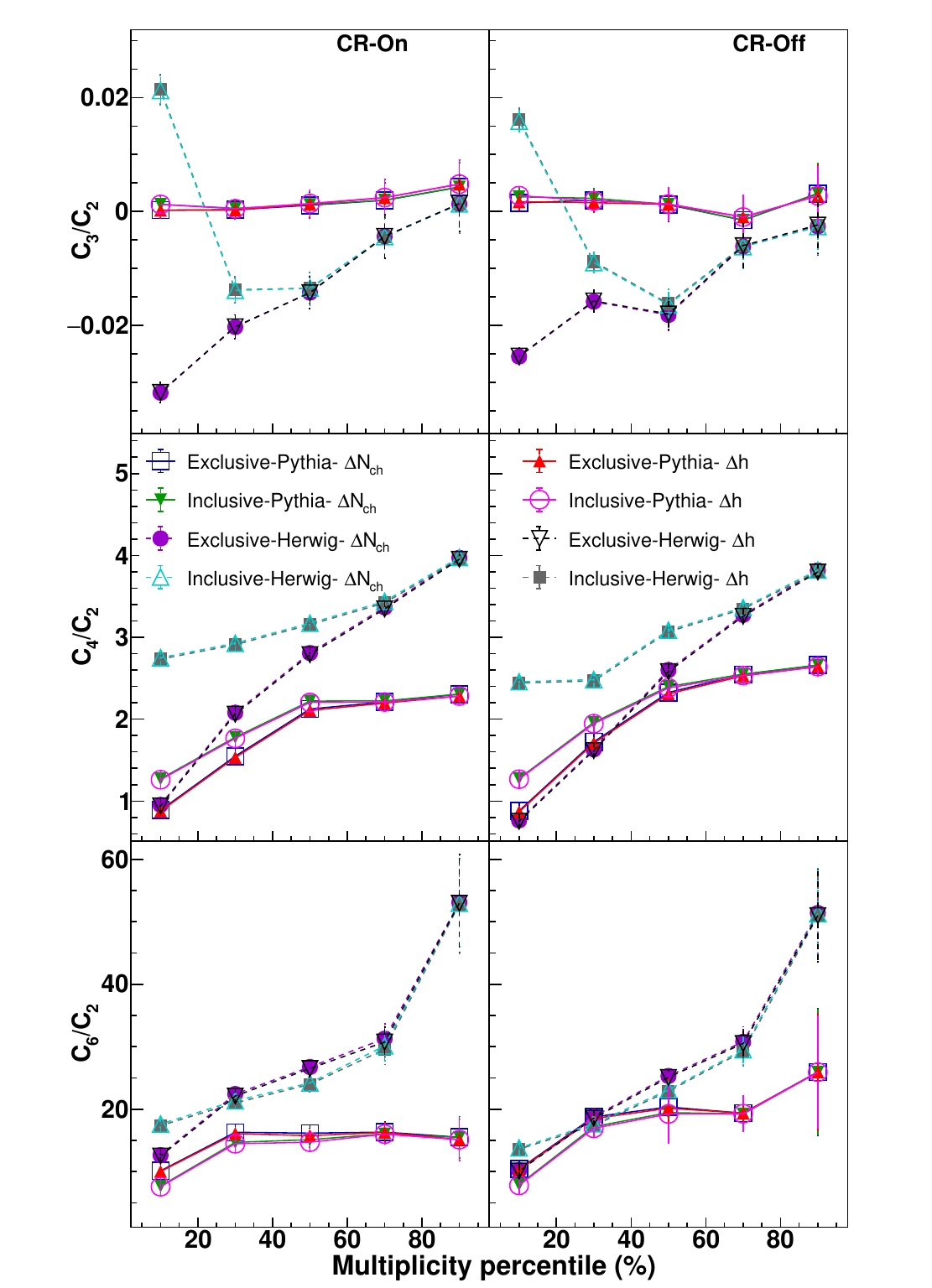}	
	\caption{Comparison of multiplicity-dependent cumulant ratios for net-charge and net-hadron distributions in \(p+p\) collisions at \( \sqrt s\) = 13 TeV, conducted using exclusive and inclusive methods. Results are presented for the 20\% multiplicity bins range. Solid lines depict Pythia8 results, and dotted lines represent Herwig results. Vertical lines denote statistical uncertainties.} 
	\label{UNC_charge}%
\end{figure}
\begin{figure}
	\centering 
	\includegraphics[width=0.49\textwidth]{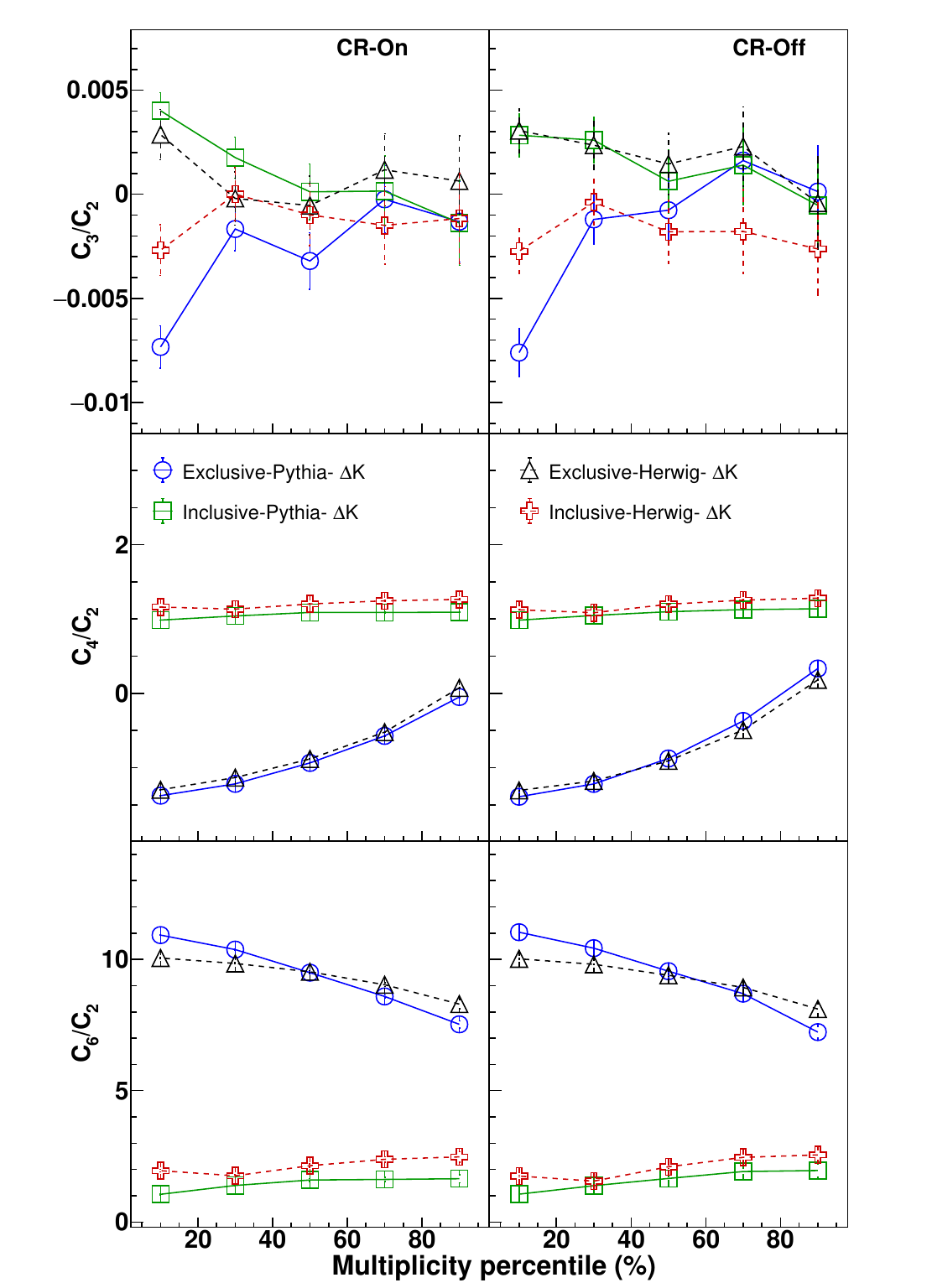}	
	\caption{Comparison of multiplicity-dependent cumulant ratios for net-kaon distributions in \(p+p\) collisions at \( \sqrt s\) = 13 TeV, conducted using exclusive and inclusive methods. Results are presented for the 20\% multiplicity bins range. Solid lines depict Pythia8 results, and dotted lines represent Herwig results. Vertical lines denote statistical uncertainties.} 
	\label{UNC_kaon}%
\end{figure}
\begin{figure}
	\centering 
	\includegraphics[width=0.49\textwidth]{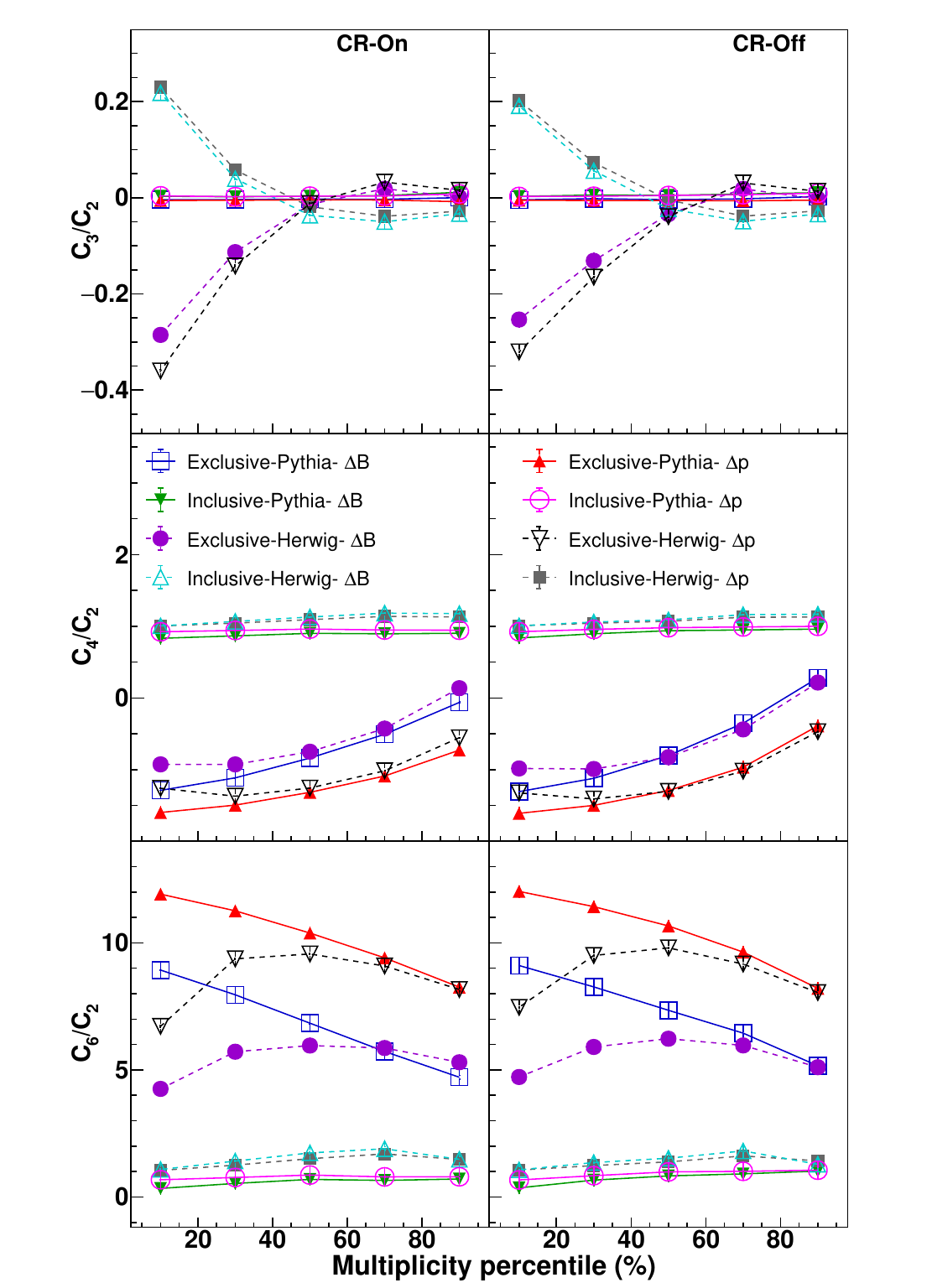}	
	\caption{Comparison of multiplicity-dependent cumulant ratios for net-baryon and net-proton distributions in \(p+p\) collisions at \( \sqrt s\) = 13 TeV, conducted using exclusive and inclusive methods. Results are presented for the 20\% multiplicity bins range. Solid lines depict Pythia8 results, and dotted lines represent Herwig results. Vertical lines denote statistical uncertainties.} 
	\label{UNC_baryon}%
\end{figure}

Figure \ref{UNC_charge} compares multiplicity-dependent cumulant ratios for net-charge and net-hadron between the exclusive and inclusive methods. Clear differences between these methods are observed across all three ratios: \(C_{3}/C_{2}, C_{4}/C_{2}\), and \(C_{6}/C_{2}\), particularly evident at lower multiplicity bins. However, these discrepancies are less pronounced at higher multiplicities. Additionally, the net-charge distribution closely mirrors that of the net-hadron.
Figure \ref{UNC_kaon} presents the multiplicity-dependent cumulant ratios obtained from exclusive and inclusive methods for net-kaon. The \(C_{3}/C_{2}\) ratio remains close to zero across all multiplicities. Notably, the \(C_{4}/C_{2}\) ratio differs significantly between the two methods: the exclusive method yields negative ratios at lower multiplicities, while the inclusive method produces values close to one across all multiplicity ranges. Conversely, substantial disparities between the exclusive and inclusive methods are observed in the \(C_{6}/C_{2}\)ratio.
Figure \ref{UNC_baryon} compares the multiplicity-dependent cumulant ratios of net-baryon and net-proton obtained from exclusive and inclusive methods. The \(C_{3}/C_{2}\) ratio shows significant differences between these methods, particularly at lower multiplicities. For the \(C_{4}/C_{2}\) ratio, negative values are obtained in the exclusive method, while the inclusive method yields positive values close to one across all multiplicity ranges. Similarly, noticeable differences between the exclusive and inclusive methods are observed for the \(C_{6}/C_{2}\) ratio, where inclusive values are close to one across all multiplicity bins.

\begin{figure}
	\centering 
	\includegraphics[width=0.49\textwidth]{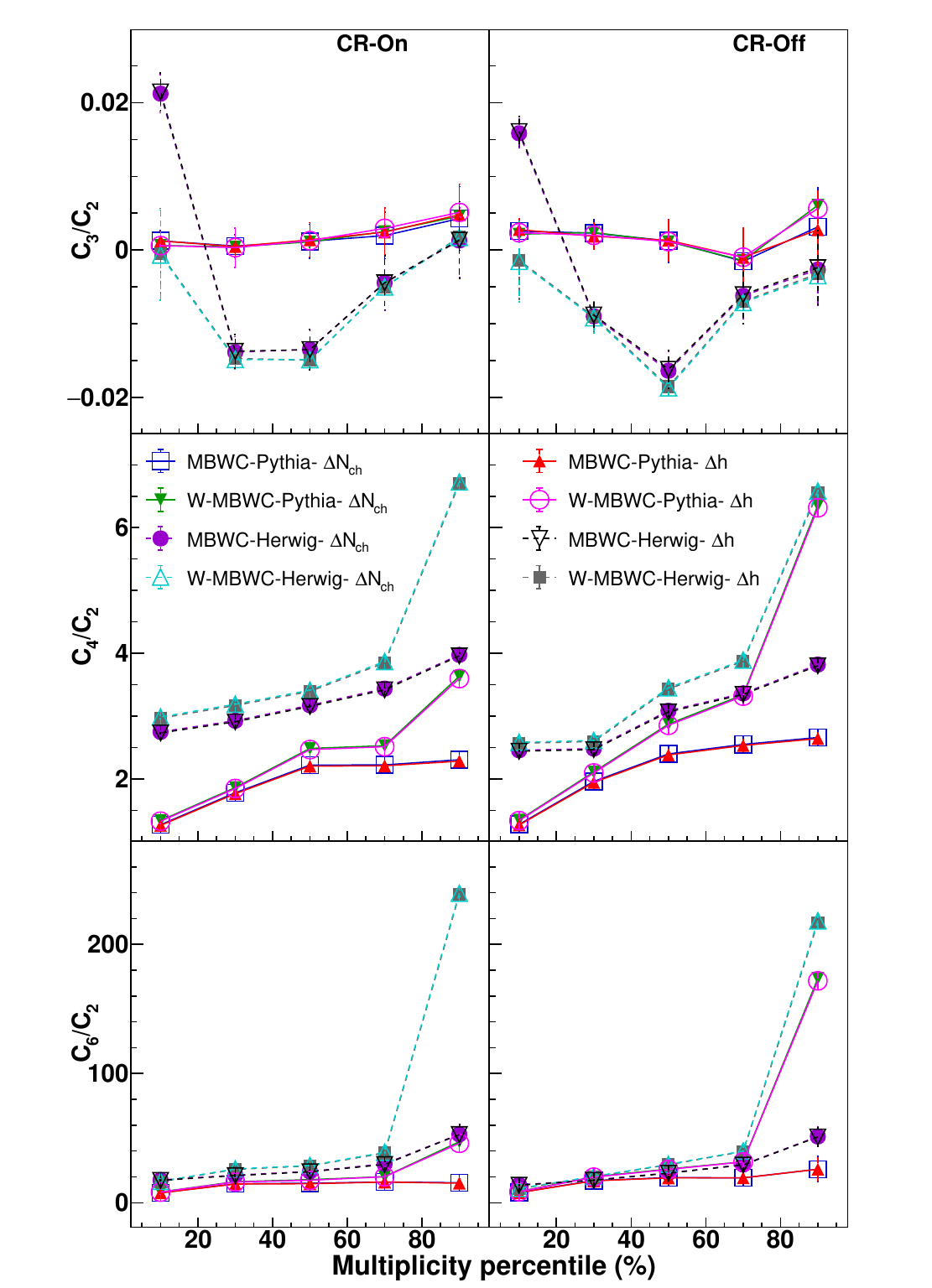}	
	\caption{Comparison of multiplicity-dependent cumulant ratios for net-charge and net-hadron distributions in \(p+p\) collisions at \( \sqrt s\) = 13 TeV, with and without MBWC method. Results are presented for the 20\% multiplicity bins range. Solid lines depict Pythia8 results, and dotted lines represent Herwig results. Vertical lines denote statistical uncertainties.} 
	\label{MBWC_charge}%
\end{figure}
\begin{figure}
	\centering 
	\includegraphics[width=0.49\textwidth]{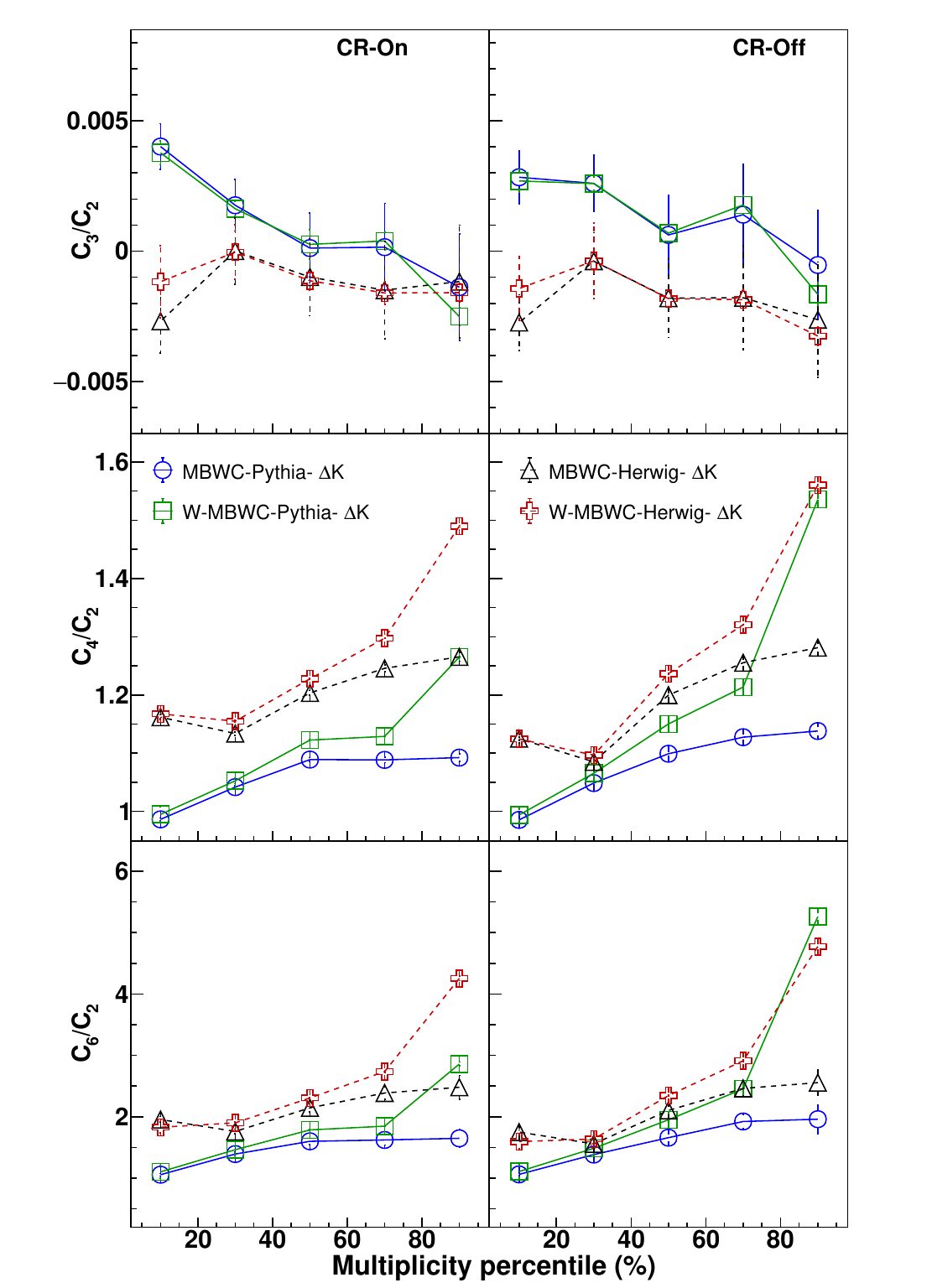}	
	\caption{Comparison of multiplicity-dependent cumulant ratios for net-kaon distributions in \(p+p\) collisions at \( \sqrt s\) = 13 TeV, with and without MBWC method. Results are presented for the 20\% multiplicity bins range. Solid lines depict Pythia8 results, and dotted lines represent Herwig results. Vertical lines denote statistical uncertainties.} 
	\label{MBWC_kaon}%
\end{figure}
\begin{figure}
	\centering 
	\includegraphics[width=0.49\textwidth]{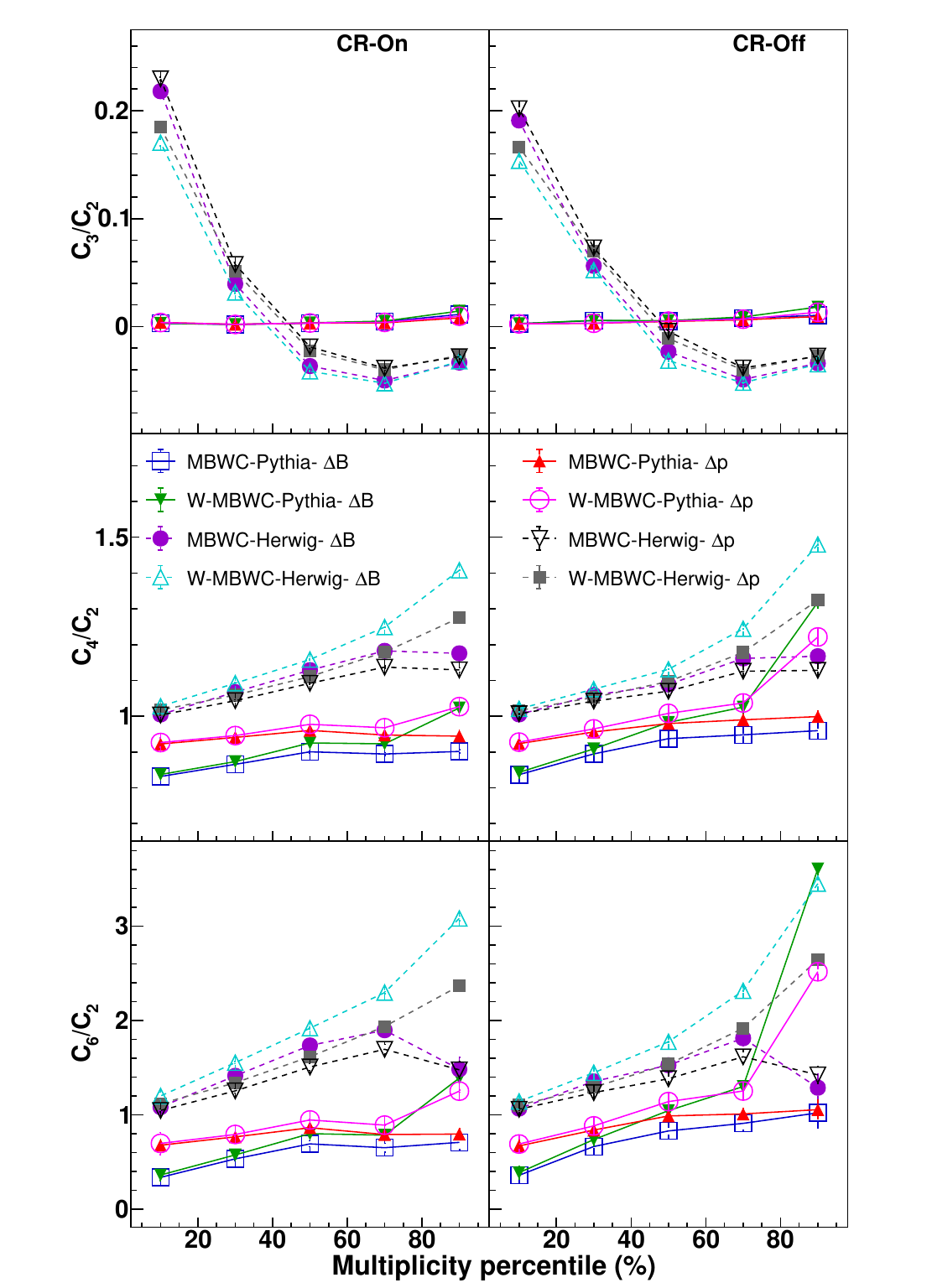}	
	\caption{Comparison of multiplicity-dependent cumulant ratios for net-baryon and net-proton distributions in \(p+p\) collisions at \( \sqrt s\) = 13 TeV, with and without MBWC method. Results are presented for the 20\% multiplicity bins range. Solid lines depict Pythia8 results, and dotted lines represent Herwig results. Vertical lines denote statistical uncertainties.} 
	\label{MBWC_baryon}%
\end{figure}
\par
Figures \ref{MBWC_charge}, \ref{MBWC_kaon}, and \ref{MBWC_baryon} illustrate the multiplicity dependence of cumulant ratios for the net-charge, net-hadron, net-kaon, net-baryon, and net-proton, with and without applying the MBWC method. In each distribution, As the order of cumulants increases, the disparities between these two methods become more apparent, especially at higher multiplicities, with values without-MBWC increase more rapidly than those with MBWC.
\par
\begin{figure}
	\centering 
	\includegraphics[width=0.49\textwidth]{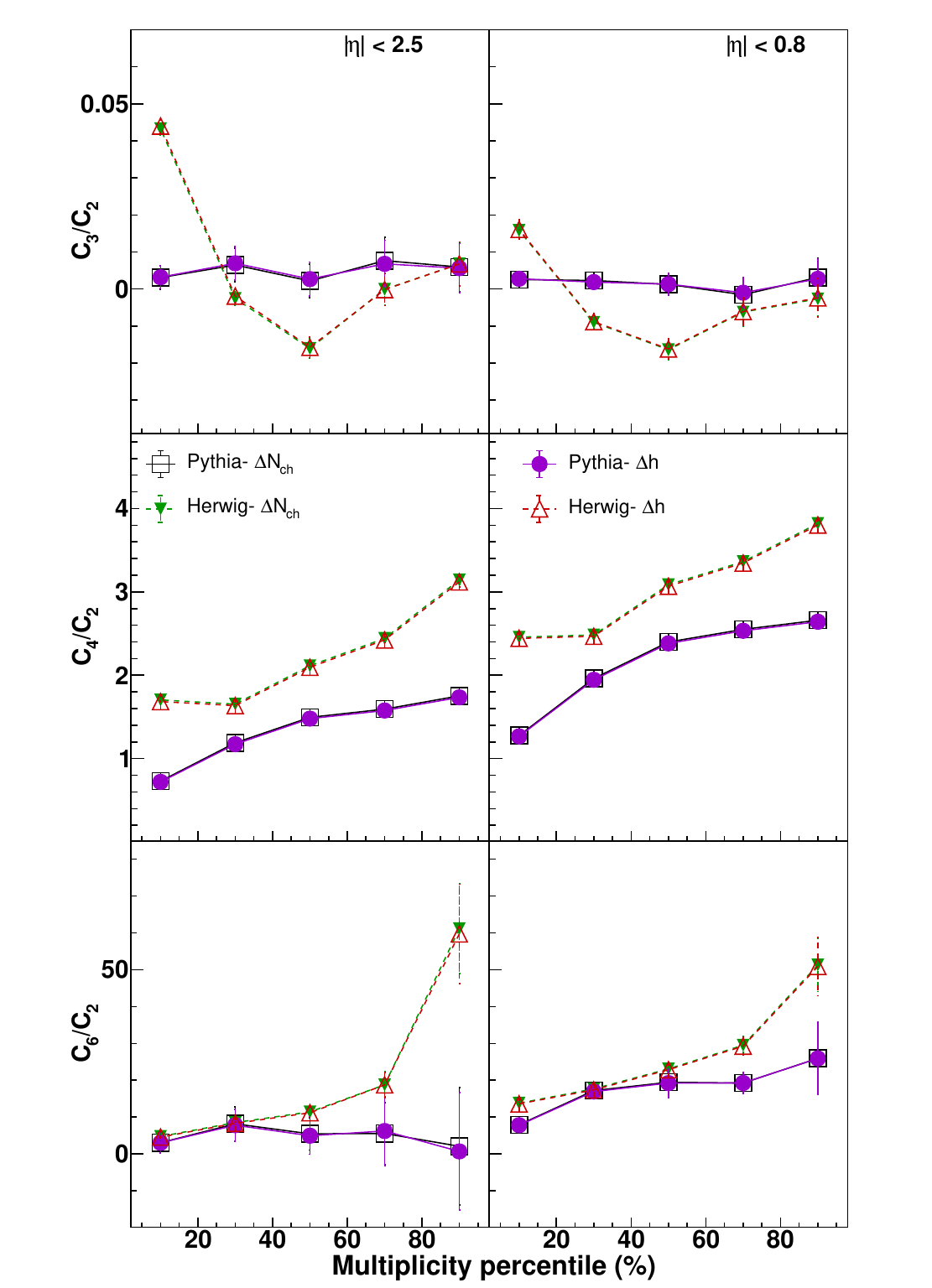}	
	\caption{Comparison of multiplicity-dependent cumulant ratios for net-charge and net-hadron distributions in \(p+p\) collisions at \( \sqrt s\) = 13 TeV under CR-off mode with different acceptance ranges. Results are presented for the 20\% multiplicity bins range. Solid lines depict Pythia8 results, and dotted lines represent Herwig results. Vertical lines denote statistical uncertainties.} 
	\label{CMS_Charge}%
\end{figure}
\begin{figure}
	\centering 
	\includegraphics[width=0.49\textwidth]{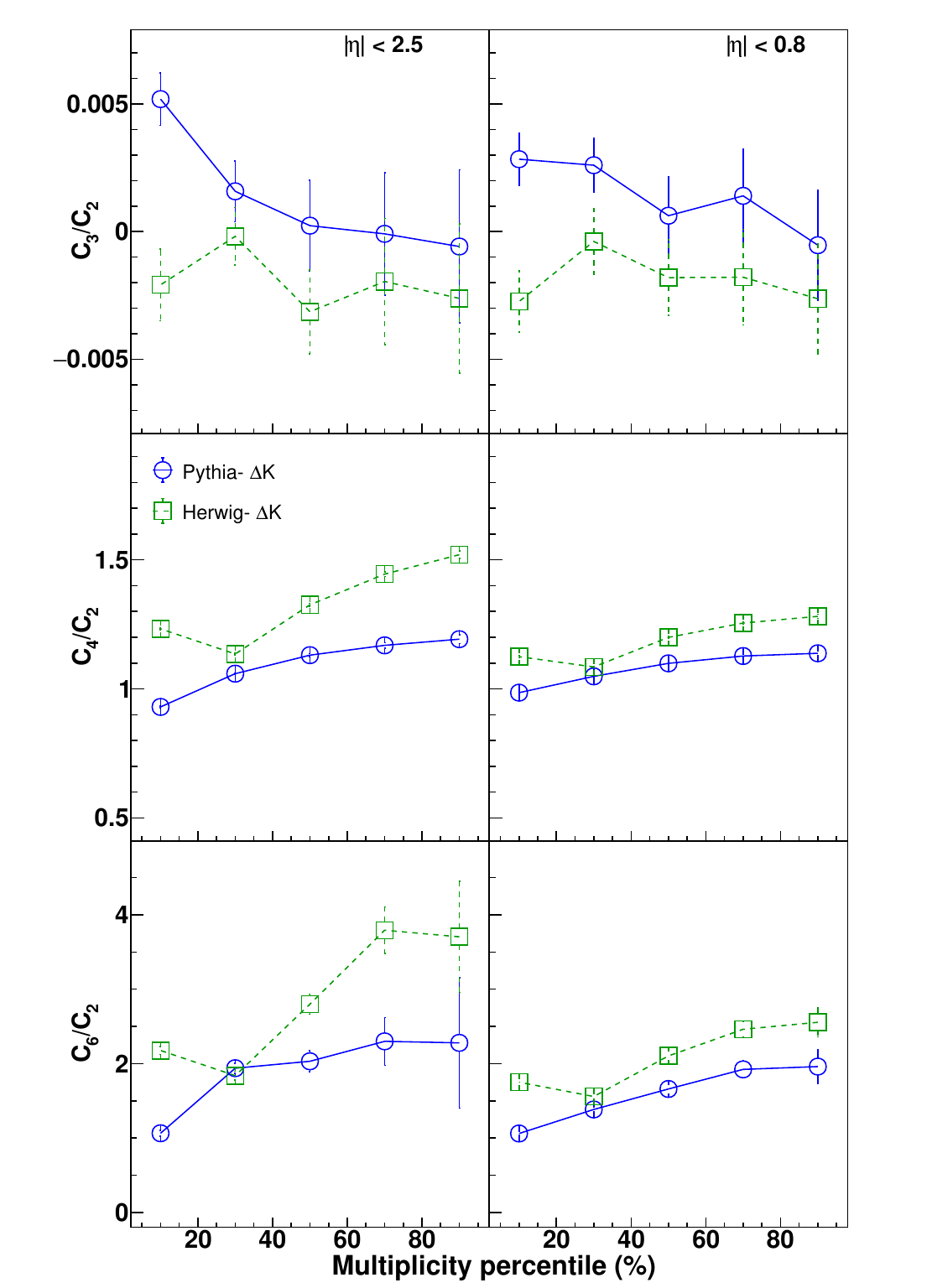}	
	\caption{Comparison of multiplicity-dependent cumulant ratios for net-kaon distributions in \(p+p\) collisions at \( \sqrt s\) = 13 TeV under CR-off mode with different acceptance ranges. Results are presented for the 20\% multiplicity bins range. Solid lines depict Pythia8 results, and dotted lines represent Herwig results. Vertical lines denote statistical uncertainties.} 
	\label{CMS_Kaon}%
\end{figure}
\begin{figure}
	\centering 
	\includegraphics[width=0.49\textwidth]{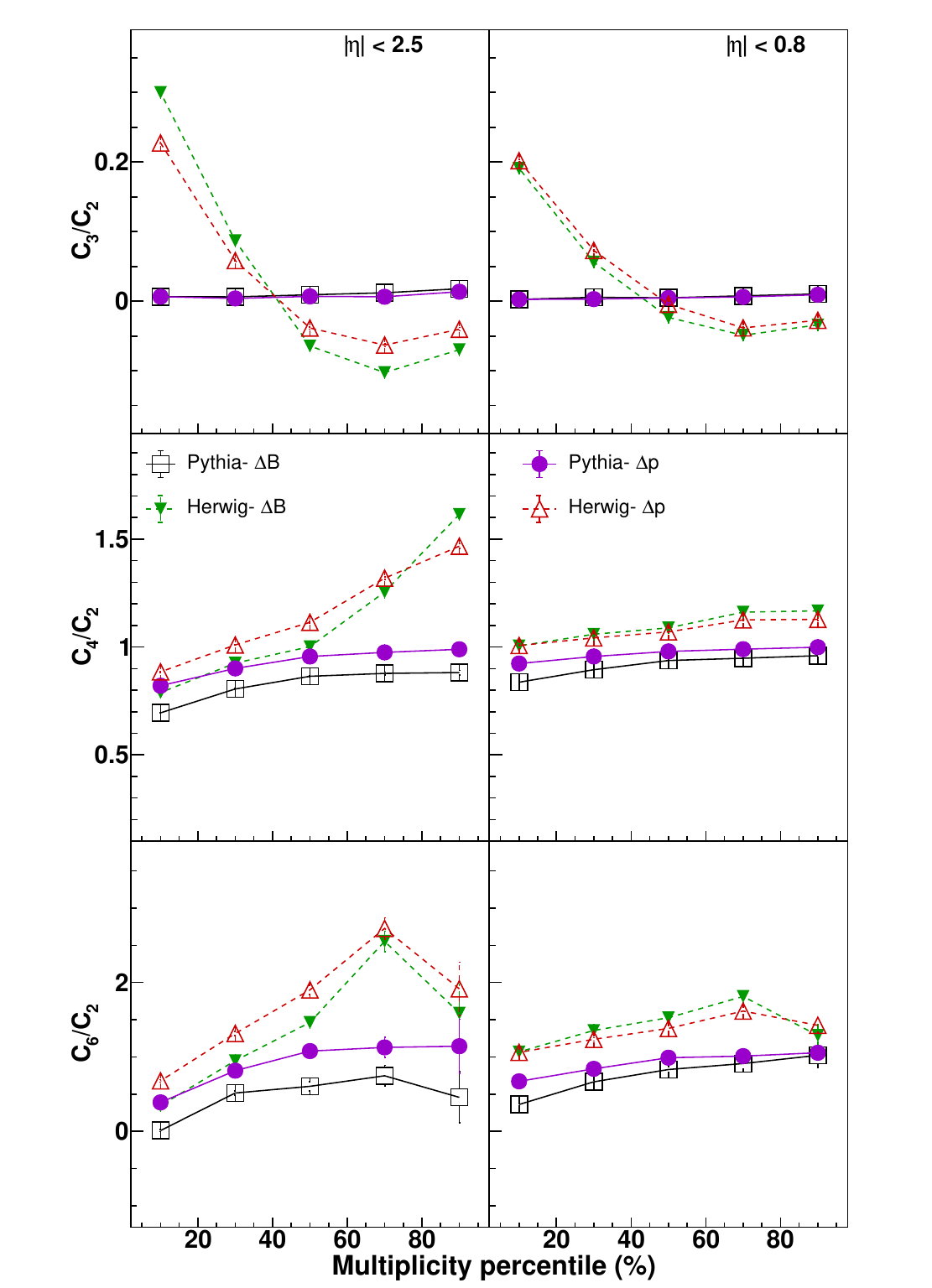}	
	\caption{Comparison of multiplicity-dependent cumulant ratios for net-baryon and net-proton distributions in \(p+p\) collisions at \( \sqrt s\) = 13 TeV under CR-off mode with different acceptance ranges. Results are presented for the 20\% multiplicity bins range. Solid lines depict Pythia8 results, and dotted lines represent Herwig results. Vertical lines denote statistical uncertainties.} 
	\label{CMS_Baryon}%
\end{figure}

Figures \ref{CMS_Charge}, \ref{CMS_Kaon}, and \ref{CMS_Baryon} illustrate the multiplicity dependence of cumulant ratios for net-charge, net-hadron, net-kaon, net-baryon, and net-proton  under the CR-off mode, considering two acceptance ranges (\(|\eta| < 2.5\) and \(|\eta| < 0.8\)). Across all distributions, there is a significant difference in yield observed when comparing the wider acceptance range to the narrower one. 

These results incorporate statistical uncertainties using 50 subgroups. Additionally, we have estimated errors using 100 subgroups, and it's worth noting that there is no significant difference between the results obtained from 50 subgroups and 100 subgroups. It's also observed that the cumulants in net-charge and net-hadron distributions exhibit higher errors than those in net-kaon, net-baryon and net-proton distributions.
\section{Summary}
We performed a detailed model calculation to understand the non-critical contributions to the cumulant observables of net-particle distributions. In this study, we examined the cumulants and cumulant ratios of net-charge, net- hadron, net-kaon, net-baryon, and net-proton multiplicity distributions in \(p+p\) collisions at $\sqrt{s}$ = 13 TeV, using Pythia8 and Herwig models. The comparison between exclusive and inclusive methods for multiplicity-dependent cumulant ratios across the net-particles reveals significant differences. Disparities are notable across different cumulants and cumulant ratios, particularly in net-kaon, net-baryon, and net- proton distributions. Additionally, the effect of MBWC on the observable was studied, which revealed increasing discrepancies in cumulant ratios with higher orders and multiplicities. This highlights the considerable influence of the MBWC method on shaping the observed cumulants of net-particle distributions.
Further analysis focused on the multiplicity dependence of cumulants up to the sixth order, revealing distinct trends for even and odd-order cumulants. Notably, negative values were observed in \(C_{4}\)  of net-kaon and net-baryon at lower multiplicities, transitioning to positive values at higher multiplicities, while net-proton consistently exhibited negative values across all multiplicities. The comparison of multiplicity-dependent cumulant ratios between exclusive and inclusive methods highlighted clear differences, particularly at lower multiplicity points, although these discrepancies were less pronounced at higher multiplicities.

\bibliography{mybib}
%

\end{document}